	\documentclass[12pt, a4paper, leqno]{article}
	
 	\usepackage{latexsym}
	\usepackage{amsmath}	
	\usepackage{epsfig}	
	\usepackage{times}	
	\usepackage{amssymb}
	\numberwithin{equation}{section}

 	\newtheorem{thr}{Theorem}[section]

	\newtheorem{lem}{Lemma}[section]
	
\begin{document}
 	\centerline{\Large{\bf Asymptotic normality of estimates}}
	\centerline{}
	\centerline{\Large{\bf in flexible seasonal time series model}}
 	\centerline{}
	\centerline{\Large{\bf with weak dependent error terms}}
 	\centerline{}
	\centerline{\textsuperscript{a}Kyong-Hui Kim, Hak-Myong Pak}
 	\centerline{}
 	\small \centerline{Faculty of Mathematics, \textbf{Kim Il Sung} University, D.P.R Korea}
	\small \centerline{\textsuperscript{a}Corresponding author. e-mail address : kim.kyonghui@yahoo.com}
	\centerline{}
	\centerline{}
	\begin{abstract}
	In this article, we consider flexible seasonal time series models which consist of a common trend function over periods 
	and additive individual trend (seasonal effect) functions.
	The consistency and asymptotic normality of the local linear estimators were obtained under the $\alpha$-mixing 
	conditions and without specifying the error distribution.  
	We develop these results to consistency and asymptotic normality of local linear estimates by using central limit 
	theorems for flexible seasonal time series model, which error terms are $k$-weak dependent and 
	$\lambda$-weak dependent random variables.
	\end{abstract}
	{\bf Keywords:} Flexible seasonal time series model; Local linear estimate; Consistency and asymptotic normality, 
	Weak dependent random variables \\
	{\bf JEL classification:} C13; C14
	%
	%
	%
	%
	\section{Introduction and previous research}
	Let $y_{t1}, y_{t2}, \cdots, ~ t=1, 2, \cdots$ are seasonal time series. The flexible model is as follows.
	\begin{equation}
	y_{tj} = T_t+S_{tj}+e_{tj}, \label{eq1.1}
	\end{equation}
	where $T_t$ is the common trend same to different periods within a season, and $S_{tj}$ is the seasonal effect, 
	satisfying $\sum_{j=1}^d S_{tj} = 0$. Semi-parametric seasonal time series model is as follows.
	\begin{equation}
	y_{tj} = \alpha(t)+\beta(t)+e_{tj}, ~ i=1, \cdots, n, ~ j=1, \cdots, d, \label{eq1.2}
	\end{equation}
	where $r_j$ is seasonal factors. Hence the overall seasonal effect changes over periods in accordance with the modulating 
	function $\beta(t)$.  Implicity, model \eqref{eq1.2} assumes that the seasonal effect curves have the same shape 
	(up to a multiplicative constant) for all seasons. We consider a more general flexible seasonal effect model having 
	the following form:
	\begin{equation}
	y_{ij} = \alpha(t_i)+\beta_j(t_i)+e_{ij}, ~ i=1, \cdots, n, ~ j=1, \cdots, d, \label{eq1.3}
	\end{equation}
	where $t_i = \frac{i}{n}, \alpha(\cdot)$ is smooth trend function in [0,1], $\{\beta_j(\cdot), j = 1, \cdots, d\}$ are 
	smooth seasonal effect functions in [0,1], either fixed or random, subject to a set of constraints, and the error term 
	$e_{ij}$ is assumed to be stationary and weak dependent random variables. As in model \eqref{eq1.2}, the following 
	constraints are needed for fixed seasonal effects:
	\begin{equation}
	\sum_{j=1}^d \beta_j(t) = 0, ~ \forall t  \label{eq1.4}
	\end{equation}
	reflecting the fact that the sum of all seasons should be zero for the seasonal factor.  In previous researches a local 
	linear technique has been used to estimate the trend and seasonal functions, and the asymptotic properties of the 
	resulting estimators have been studied assuming that error terms were $\alpha$-mixing random variables \cite{cai1}. 
	Also asymptotic properties of nonparametric estimators for various time series models has been studied by local 
	linear method \cite{cai2, cai3, cai4, kim}.\\

	\noindent \textbf{Weak dependence and Problems}

	In model \eqref{eq1.3}, statistical properties of weighted least square estimators are depended conclusively on statistical 
	structure of dependent error terms. Many authors have used the two type of dependence: one is, mixing properties 
	introduced by Rosenblatt(1956); another is, martingales approximations or mixingales, following the works of 
	Gordin(1969, 1973) and Mc Leisch(1974, 1975). 
	Concerning strongly mixing sequences, very deep and elegant results have been established by Rio(2000) and Bradley(2002).
	However, many classes of time series do not satisfy any mixing condition, conversely most of such time series enter 
	the scope of mixingales but limit theorems and moment inequalities are more difficult to obtain in this general setting, 
	so between those directions Bickel and B\"uhlmann(1999) and seperatively Doukhan and Louhichi(1999) introduced a new 
	idea of weak dependence. Their concept of weak dependence makes explicit the asymptotic independence between 
	\textquoteleft past\textquoteright and \textquoteleft future\textquoteright: this means that the \textquoteleft past\textquoteright 
	is progressively forgotten. Roughly speaking, for convenient functions $f$ and $g$, they assumed that
	\begin{equation*}
	\text{Cov}(f(\text{\textquoteleft} \text{past}\text{\textquoteright)}, g(\text{\textquoteleft} \text{future}\text{\textquoteright}))
	\end{equation*}
	is small when the distance between the \textquoteleft past\textquoteright and the \textquoteleft future\textquoteright is 
	sufficiently large. The main advantage is that such a kind of dependence contains lots of pertinent examples and can be 
	used in various situations. Therfore the central limit theorems for weak dependent variables has been studied in recent 
	years \cite{bar, bul, ded}. In this article, we are going to derive consistency and asymptotic normality of the weighted 
	least square estimators with a local linear method, assuming that error terms are $k$-weak dependent and $\lambda$-weak 
	dependent random variables.

	%
	%
	%
	%

	\section{Main results and proof of theorems}
	Combination of \eqref{eq1.3} and \eqref{eq1.4} in a matrix expression leads to $\theta$.
	\begin{equation}
	\mathbf{Y}_i = \mathbf{A}\theta(t_i)+\mathbf{e}_i \label{eq2.1}
	\end{equation}
	where
	\begin{equation*}
	\mathbf{Y}_i = \left(
	\begin{array}{c}
		y_{i1} \\
		\vdots \\
		y_{id}
	\end{array} \right),
	\mathbf{A} = \left(
	\begin{array}{c c}
		\mathbf{1}_{d-1} & \mathbf{I}_{d-1} \\
		1 & -\mathbf{1}_{d-1}
	\end{array} \right),
	\theta(t) = \left(
	\begin{array}{c}
		\alpha(t) \\
		\beta_1(t) \\
		\vdots \\
		\beta_{d-1}(t)
	\end{array} \right),
	\mathbf{e}_i = \left(
	\begin{array}{c}
		e_{i1} \\
		e_{i2} \\
		\vdots \\
		e_{id}
	\end{array} \right).
	\end{equation*}
	$\mathbf{I}_d$ is the $d \times d$ identity matrix, and the error term $\mathbf{e}_i$ is assumed to be stationary 
	with $E(\mathbf{e}_i) = 0$ and $\text{cov}(\mathbf{e}_i, \mathbf{e}_j) = \mathbf{R}(i − j)$. 
	Assuming that $\alpha(\cdot)$ and $\beta_j(\cdot)$ have a continuous second derivative in [0, 1], then 
	$\alpha(\cdot)$ and $\beta_j(\cdot)$ can be approximated by linear functions at any time point $0 \leq t \leq 1$ as follows:
	\begin{equation*}
	\left\{
	\begin{array}{l}
	\alpha(t_i) \cong a_0+b_0(t_i-t)  \\
	\beta_j(t_i) \cong a_j+b_j(t_i-t), ~ 1 \leq j \leq d-1,
	\end{array} 
	\right.
	\end{equation*}
	where $\cong$ denotes the first order Taylor approximation. Hence $\theta(t_i) \cong \mathbf{a}+\mathbf{b}(t_i-t)$, 
	where $\mathbf{a} = \theta(t)$ and $\mathbf{b} = \theta^{(1)}(t) = d\theta(t)/dt$ and \eqref{eq2.1} is approximated by
	\begin{equation*}
	\mathbf{Y}_i \cong \mathbf{Z}_i \left(
	\begin{array}{c}
	\mathbf{a}  \\
	\mathbf{b}
	\end{array} 
	\right) + \mathbf{e}_i,
	\end{equation*}
	where $\mathbf{Z}_i = (\mathbf{A}, (t_i - t)\mathbf{A})$. Therefore, the locally weighted sum of least squares is
	\begin{equation}
	\sum_{i=1}^n \left\{ \mathbf{Y}_i - \mathbf{Z}_i \left(
	\begin{array}{c}
	\mathbf{a}  \\
	\mathbf{b}
	\end{array}
	\right) \right\} \left\{ \mathbf{Y}_i - \mathbf{Z}_i \left(
	\begin{array}{c}
	\mathbf{a}  \\
	\mathbf{b}
	\end{array}
	\right) \right\} K_h(t_i-t), \label{eq2.2}
	\end{equation}
	where $K_h(u) = K(u/h)/h, K(\cdot)$ is the kernel function, and $h = h_n > 0$ is the bandwidth satisfying $h \to 0$ 
	and $nh \to \infty$ as $n \to \infty$, which controls the amount of smoothing used in the estimation.
	By minimizing \eqref{eq2.2} with respect to $\mathbf{a}$ and $\mathbf{b}$, we obtain the local linear estimate 
	$\hat{\theta}(t) = \hat{\mathbf{a}}, \hat{\theta}'(t) = \mathbf{b}'$.\\

	%
	%

	\noindent \textbf{Assumptions:}

	$A1$. Assume that the kernel $K(u)$ is symmetric and satisfies the Lipschitz condition and $uK(u)$ is bounded, 
	and that $\alpha(\cdot)$ and $\beta_j(\cdot)$ have continuous second derivatives in [0, 1].

	$A2$. For each $n, \{ \mathbf{e}_{n1}, \cdots, \mathbf{e}_{nn}\}$ have the same joint distribution as 
	$\{ \xi_1, \xi_2, \cdots, \xi_n\}$, where $\xi_t, t = \cdots, -1, 0, 1, \cdots$ is a strictly stationary time series with 
	the covariance matrix $\mathbf{R}(k - l) = \text{cov}(\xi_k, \xi_l)$. 
	Assume that the time series $\{\xi_t\}$ is sequence of $k$-weak dependent random vectors with the finite $(2 + \zeta)$th 
	moment for some $\zeta > 0$ (i.e. $E \|\xi_i\|^{2+\zeta}<\infty)$ and $k$-weak dependent coefficient 
	$K_{\mathbf{e}}(r)$ satisfying $K_{\mathbf{e}}(r) = 0(h^{-2}r^{-k})$, where $k > 2 + 1/\zeta$.

	$\bar{A}2$. Assume that the time series $\{ \xi_t\}$ is sequence of $\lambda$-weak dependent random vectors 
	satisfying the assumption $A2$ and $\lambda_{\mathbf{e}}(r) = 0(h^{-2}r^{-\lambda})$, where $\lambda > 4 + 2/\zeta$.\\

	%
	%

	\noindent \textbf{Main results:}

	%
	%

	\begin{lem}\label{lem2.1}
	Let a sequence of random vectors $\{\mathbf{e}_k\}$ is stationary with mean 0 and $k$-weak dependent ($\lambda$-
	weak dependent) and $\{ z_k\}$ is a sequence of stationary random variables defined as follows:
	\begin{equation*}
	z_k = hK_h(t_i-t) \mathbf{d}'\,\mathbf{e}_k,
	\end{equation*}
	then $\{ z_k\}$ are also $k$-weak dependent ($\lambda$-weak dependent) sequence and the following equality holds.
	\begin{equation*}
	|h|^2K_{\mathbf{e}}(r) = K_z(r), ~ |h|^2 \lambda_{\mathbf{e}}(r) = \lambda_z(r),
	\end{equation*}
	where $K_{\mathbf{e}}(r), K_z(r)$ and $\lambda_{\mathbf{e}}(r), \lambda_z(r)$ are $k$-weak dependent and 
	$\lambda$-weak dependent coefficients respectively of $\{\mathbf{e}_k\}, \{z_k\}$.
	\end{lem}
	\emph{Proof}.
	\begin{equation*}
	K_z(r) = \underset{u,v}{\text{sup}} \underset{(i,j)\in \Gamma(u,v,r)}{\text{sup}} \underset{f\in \Im_u g \in \jmath_v}
	{\text{sup}} \frac{\left\vert\text{cov}\Big( f(z_{i1}, \cdots, z_{iu}), g(z_{j1}, \cdots, z_{ju}) \Big) \right\vert}{\psi(f, g)}
	\end{equation*}
	where in case of $k$-weak dependence $\Im_u (\Im_u = \jmath_u)$ is the wider set of functions from $\chi^u$ to 
	$\mathbf{R}$, which are Lipschitz with respect to the distance $\delta_1$ on $\chi^u$ defined by
	\begin{equation*}
	\delta_1(x, y) = \sum_{i=1}^u \delta(x_i, y_i),
	\end{equation*}
	but which are not necessarily bounded.
	In this case
	\begin{equation*}
	\psi(f, g) = d_f d_g Lip(f) Lip(g)
	\end{equation*}
	and in case of $\lambda$-weak dependence $\Im_u (\Im_u = \jmath_u)$ is the set of bounded functions from $\chi^u$ 
	to $\mathbf{R}$, which are Lipschitz with respect to the distance $\delta_1$ on $\chi^u$ defined by same method,
	\begin{equation*}
	\psi(f, g) = d_f \| g\|_{\infty} Lip(f) + d_g \| f\|_{\infty} Lip(g) + d_fd_g Lip(f) Lip(g).
	\end{equation*}
	And then $\delta(x_i, y_i)$ is a distance on a space $\chi$, in case of $\{ z_k\}$ we have $\chi = \mathbf{R}$ and 
	$\delta(x_i, y_i) = |x_i-y_i|$.
 	Now we define $\tilde{\delta}(\mathbf{e}_i, \mathbf{e}_j)$ on $\mathbf{R}^d$ by
	\begin{equation*}
	\tilde{\delta}(\mathbf{e}_i, \mathbf{e}_j) = \delta \Big( K_h(t_i-t)\mathbf{e}_i, K_h(t_j-t)\mathbf{e}_j\Big),
	\end{equation*}
	where $\delta$ is a usual distance defined by $\| \cdot \|$ on $\chi = \mathbf{R}^d$. We define
	\begin{equation*}
	F(\mathbf{e}_{i1}, \cdots, \mathbf{e}_{iu}) = f(z_{i1}, \cdots, z_{iu}).
	\end{equation*}
	Then the following relations hold:
	\begin{align*}
	& F(\mathbf{e}_{i1}, \cdots, \mathbf{e}_{iu}) - F(\mathbf{e}_{k1}, \cdots, \mathbf{e}_{ku})  \\
	& \qquad \leq Lip(f) \sum_{l=1}^u \vert h\mathbf{d}'(K_h(t_{il}-t) \mathbf{e}_{il} - K_h(t_{kl}-t) \mathbf{e}_{kl})\vert \\
	& \qquad \leq Lip(f) \| \mathbf{d} \| \vert h \vert \sum_{l=1}^u \| (K_h(t_{il}-t) \mathbf{e}_{il} - K_h(t_{kl}-t) \mathbf{e}_{kl}) \| \\
	& \qquad = Lip(f) \| \mathbf{d} \| \vert h \vert \sum_{l=1}^u \delta(K_h(t_{il}-t) \mathbf{e}_{il} - K_h(t_{kl}-t) \mathbf{e}_{kl})  \displaybreak[0] \\
	& \qquad = Lip(f) \| \mathbf{d} \| \vert h \vert \tilde{\delta}_1 \Big( (\mathbf{e}_{i1}, \cdots, \mathbf{e}_{iu}), 
	(\mathbf{e}_{k1}, \cdots, \mathbf{e}_{ku}) \Big).
	\end{align*}
	Therefore, Lipschitz constant of $F$ is
	\begin{align*}
	& Lip(F) = Lip(f) \|d \| \vert h\vert = Lip(f)\vert h \vert, \\
	& Lip(G) = Lip(g) \vert h\vert, ~ d_f = d_F, ~ d_g = d_G,
	\end{align*}
	so
	\begin{equation*}
	\psi(f, g) = \vert h \vert^{-2} \psi(F, G).
	\end{equation*}
	Hence
	\begin{eqnarray*}
	K_z(r) & = & \underset{u,v}{\text{sup}} \underset{(i,j)\in \Gamma(u,v,r)}{\text{sup}} \underset{f\in \Im_u g \in \jmath_v}
	{\text{sup}} \frac{\left\vert\text{cov}\Big( f(z_{i1}, \cdots, z_{iu}), g(z_{j1}, \cdots, z_{ju}) \Big) \right\vert}{\psi(f, g)} \\
	& = & \underset{u,v}{\text{sup}} \underset{(i,j)\in \Gamma(u,v,r)}{\text{sup}} \underset{f\in \tilde{\Im}_u g 
	\in \tilde{\jmath}_v}{\text{sup}} \frac{\left\vert\text{cov}\Big( F(\mathbf{e}_{i1}, \cdots, \mathbf{e}_{iu}),
	 G(\mathbf{e}_{i1}, \cdots, \mathbf{e}_{iu}) \Big) \right\vert}{\psi(f, g)} \vert h \vert^2 \\
	& = & K_{\mathbf{e}}(r) \vert h \vert^2.
	\end{eqnarray*}

	Finally convergence of two weak dependent coefficients are equivalent, hence $\{z_k\}$ are also $k$-weak 
	dependent sequence. Correspondingly equivalence of $\lambda$-weak dependence is proved. $\Box$

	%
	%

	\begin{lem}\label{lem2.2}
	Under assumptions of Lemma~\ref{lem2.1}, we have
	\begin{equation*}
	\lim_{n \to \infty} \mathbf{DB}_{n0} = v_0 \Sigma_0, ~~ \mathbf{B}_{n1} \overset{p}{\rightarrow} 0,
	\end{equation*}
	where for $k = 0, 1,$
	\begin{equation*}
	\mathbf{B}_{nk} = (h/n)^{1/2} \sum_{i=1}^n (t_i-t)^k \mathbf{e}_{ni}K_h(t_i-t), ~ k=1, 2.
	\end{equation*}
	\end{lem}
	\emph{Proof}. By the stationarity of $\{\xi_j\}$,
	\begin{align*}
	&\mathbf{DB}_{n0} = n^{-1}h \sum_{1 \leq k, l \leq n}\mathbf{R}(k-l)K_h(t_k-t)K_h(t_l-t) \\
	& \qquad = n^{-1}h \mathbf{R}(0) \sum_{k=1}^n K_h^2(t_k-t)+2n^{-1}h \sum_{1 \leq l <k \leq n}\mathbf{R}(k-l)K_h(t_k-t)K_h(t_l-t) \\
	& \qquad =: D_1+D_2.
	\end{align*}
	Clearly, by the Riemann sum approximation of an integral,
	\begin{equation*}
	D_1 \approx \mathbf{R}(0)h \int_0^1K_h^2(u-t)du \approx v_0 \mathbf{R}(0).
	\end{equation*}
	Since $nh \to \infty$, there exists $c_n \to \infty$ such that $c_n/(nh) \to 0$. Let $S_1 = \{ (k, l) : 1 \leq k-l \leq c_n ; 
	1 \leq l<k \leq n \}$ and $S_2 = \{ (k, l) : 1 \leq l <k \leq n \} \setminus S_1$. Then, $D_2$ is split unto two terms as 
	$\sum_{S_1}(\cdots)$, donoted by $D_{21}$ and $\sum_{S_2}(\cdots)$, donoted by $D_{22}$. By assumptions of 
	Lemma~\ref{lem2.1}, we have
	\begin{eqnarray*}
	\vert D_{22(jm)} \vert & \leq & Cn^{-1}h \sum_{S_2} \vert r_{jm}(k-l) \vert K_h(t_k-t)K_h(t_l-t) \\
	& \leq & Cn^{-1}h \sum_{S_2} K(k-l)K_h(t_k-t)K_h(t_l-t) \\
	& \leq & Cn^{-1} \sum_{k=1}^n K_h(t_k-t) \sum_{k_1>A_n}K(k_1) \\
	& \leq & C\sum_{k_1>A_n}k_1^{-(2+1/\zeta)} \\
	& \leq & CA_n^{-1/\zeta}\sum_{k_1>A_n}k_1^{-2}.
	\end{eqnarray*}
	Since $A_n \to \infty$, the right side of above expression converges to zero. For any $(k, l) \in S_1$, by Assumption $A1$
 	\begin{equation*}
	\vert K_h(t_k-t) - K_h(t_l-t) \vert \leq Ch^{-1}(t_k-t_l)/h \leq CA_n/(nh^2).
	\end{equation*}
	From this inequality and the result of Lemma 4.2 in \cite{ded},
	\begin{eqnarray*}
	\vert I \vert & = & \left\vert 2n^{-1}h \sum_{l=1}^{n-1} \sum_{1 \leq k-l \leq A_n} r_{jm}(k-l)\left\{K_h(t_k-t)-K_h(t_l-t)\right\}K_h(t_l-t) \right\vert \\
	& \leq & CA_nn^{-2}h^{-1} \sum_{l=1}^{n-1} \sum_{1 \leq k-l \leq A_n} \vert r_{jm}(k-l)\vert K_h(t_l-t) \\
	& \leq & CA_nn^{-2}h^{-1} \sum_{l=1}^{n-1} K_h(t_l-t) \sum_{k \geq 1} \vert r_{jm}(k)\vert \\
	& \leq & CA_n/(nk) \to 0.
	\end{eqnarray*}
	Also the following result hold
	\begin{eqnarray*}
	D_{21} & = & 2n^{-1}h \sum_{l=1}^{n-1} \sum_{1 \leq k-l \leq A_n} r_{jm}(k-l) K_h(t_k-t)K_h(t_l-t) \\
	& = & 2n^{-1}h \sum_{l=1}^{n-1} K_h^2(t_l-t) \sum_{1 \leq k-l \leq A_n} r_{jm}(k-l)+I.
	\end{eqnarray*}
	Therefore
	\begin{equation*}
	\lim_{n \to \infty}D_{21} = 2v_0 \sum_{k=1}^{\infty}r_{jm}(k),
	\end{equation*}
	hence
	\begin{equation*}
	\lim_{n \to \infty} \mathbf{D}B_{n0} = v_0 \left[ \mathbf{R}_0 + 2\sum_{k=1}^{\infty} \mathbf{R}(k)\right] = v_0\Sigma_0.
	\end{equation*}
	Otherwise, by the assumption A1, we get the following
	\begin{equation*}
	\mathbf{D}B_{n1} = n^{-1}h \sum_{1 \leq k, l \leq n} \mathbf{R}(k-l)(t_k-t)(t_l-t)K_h(t_k-t)K_h(t_l-t)
	\end{equation*}
	and
	\begin{equation*}
	Cn^{-1}h \sum_{1 \leq k, l \leq n} \left\vert \mathbf{R}(k-l) \right\vert \leq Chn^{-1}\sum_{k=-\infty}^{\infty} \vert \mathbf{R}(k)\vert \to 0. \qquad \Box
	\end{equation*}
	
	%
	%

	\begin{thr}\label{thr2.1}
	 Under Assumptions $A1$ and $A2$, (or $A1, \bar{A}2$, we have
	\begin{equation*}
	\hat{\theta}(t)-\theta{t}-\frac{h^2}{2}\mu_2\theta^{(2)}(t)+o(h^2) = O_p\left( (nh)^{-1/2}\right).
	\end{equation*}
	\end{thr}
	\emph{Proof}. Let $\mu_k = \int u^k K(u)du, ~ v_k = \int u^k K^2(u)du$, then
	\begin{equation}
	\lim_{n \to \infty} S_{n,k}(t) = h^k\mu_k \label{eq2.3}
	\end{equation}
	From Taylor explanation, we have
	\begin{equation*}
	\theta(t_i) = \theta{t}+\theta'(t)(t_i-t)+\frac{\theta^{(2)}(t)}{2!}(t_i-t)^2+o(h^2),
	\end{equation*}
	hence it follows that
	\begin{eqnarray*}
	n^{-1}\sum_{i=1}^n (t_i-t)^k \theta(t_i)K_h(t_i-t) &=& S_{n,k}(t)\theta(t) + S_{n,k+1}(t)\theta'(t) \\
		& & \quad +\frac{1}{2}S_{n,k+2}(t)\theta^{(2)}(t)+o(h^2).
	\end{eqnarray*}
	By the model \eqref{eq2.1}
	\begin{equation*}
	\mathbf{Y}_i = A\theta(t_i)+\mathbf{e}_i = A\left(\theta{t}+\theta'(t)(t_i-t)+\frac{\theta^{(2)}(t)}{2!}(t_i-t)^2+o(h^2) \right) + \mathbf{e}_i
	\end{equation*}
	and applying the least square estimation result of \cite{cai1}
	\begin{eqnarray*}
	\hat{\theta}(t) & = & A^{-1} \sum_{i=1}^n S_i(t) \mathbf{Y}_i = \theta(t) + \frac{1}{2}\frac{S_{n,2}^2(t)-S_{n,1}(t)S_{n,3}(t)}{S_{n,0}(t)S_{n,2}(t)-S_{n,1}^2(t)} \theta^{(2)}(t)\\
		& & \quad + o(h^2) + A^{-1}\sum_{i=1}^n S_i(t)\mathbf{e}_{ni}.
	\end{eqnarray*}
	Then, by assumption $A1$ and using that $\mu_1=0, \mu_3=0, \mu_0=1$
	\begin{equation*}
	\hat{\theta}(t)-\theta(t)-\frac{h^2}{2}\mu_2\theta^{(2)}(t) + o(h^2) = A^{-1}\sum_{i=1}^n S_i(t) \mathbf{e}_{ni}, 
	\end{equation*}
	which implies that
	\begin{equation}
	\sqrt{nh}\left\{ \hat{\theta}(t)-\theta(t)-\frac{h^2}{2}\mu_2 \theta^{(2)}(t)+o(h^2) \right\} = A^{-1} \frac{S_{n,2}(t)\mathbf{B}_{n0}-
	S_{n,1}(t)\mathbf{B}_{n1}}{S_{n,0}(t)S_{n,2}(t)-S_{n,1}^2(t)}, \label{eq2.4}
	\end{equation}
	where both $\mathbf{B}_{n0}$ and $\mathbf{B}_{n1}$ are defined in Lemma~\ref{lem2.2}.  From Lemma~\ref{lem2.2} and 
	Eq.\eqref{eq2.3}, we prove the theorem. $\Box$
	
	%
	%

	\begin{thr}\label{thr2.2}
	 Under Assumptions $A1$ and $A2$, (or $A1, \bar{A}2$, we have
	\begin{equation*}
	\sqrt{nh}\left\{ \hat{\theta}(t)-\theta(t)-\frac{h^2}{2}\mu_2 \theta^{(2)}(t)+o(h^2) \right\} \to N(0, \Sigma_{\theta}),
	\end{equation*}
	where $\Sigma_{\theta} = v_0A^{-1}\Sigma_0(A^{-1})'$.
	\end{thr}
	\emph{Proof}. From Eq.\eqref{eq2.4}, we get
	\begin{eqnarray*}
	&& \sqrt{nh}\left\{ \hat{\theta}(t)-\theta(t)-\frac{h^2}{2}\mu_2 \theta^{(2)}(t)+o(h^2) \right\}  \\
	&& \qquad \qquad = A^{-1} \frac{S_{n,2}(t)}{S_{n,0}(t)S_{n,2}(t)-S_{n,1}^2(t)} \left\{ \mathbf{B}_{n0}-\frac{S_{n,1}(t)}{S_{n,2}(t)}\mathbf{B}_{n1} \right\}.
	\end{eqnarray*}
	So
	\begin{eqnarray*}
	&& \frac{S_{n,1}(t)}{S_{n,2}(t)}\mathbf{B}_{n1} = \left\{ \frac{S_{n,1}(t)-\mu_1h}{S_{n,2}(t)} + \frac{\mu_1h}{S_{n,2}(t)}\right\} \mathbf{B}_{n1}, \\
	&& \frac{S_{n,2}(t)}{S_{n,0}(t)S_{n,2}(t)-S_{n,1}^2(t)}= 1+\frac{S_{n,1}^2(t)-S_{n,2}(t)S_{n,0}(t)+S_{n,2}(t)}{S_{n,0}(t)S_{n,2}(t)-S_{n,1}^2(t)}.
	\end{eqnarray*}
	By Lemma~\ref{lem2.2} and Assumption $A1$, we have
	\begin{equation*}
	\frac{S_{n,1}(t)}{S_{n,2}(t)}\mathbf{B}_{n1} \overset{p}{\to} 0, \quad \frac{S_{n,2}(t)}{S_{n,0}(t)S_{n,2}(t)-S_{n,1}^2(t)} \to 1,
	\end{equation*}
	which implies that to establish the asymptotic normality of $\hat{\theta(t)}$, we only need to consider the asymptotic normality for $\mathbf{B}_{n0}$.
	
	Hence it remains to prove the asymptotic normality of $\mathbf{d'B}_{n0}$ for all $d \in \mathbf{R}^d (\|d \|=1)$. Let $Z_{ni} = 
	\sqrt{h}\mathbf{d'e}_{ni} K_h(t_i − t)$, then clearly $\mathbf{d'B}_{n0} = \frac{1}{\sqrt{n}}\sum_{i=1}^nZ_{ni}$. Moreover
	\begin{equation}
	\mathbf{D}(\mathbf{d'B}_{n0}) = v_0\mathbf{d'}\Sigma_0 \mathbf{d} \{1+o(1)\} = \theta_d^2\{1+o(1)\}
	\end{equation}
	Since $k$-weak dependence and $\lambda$-weak dependence of $\{Z_{ni}\}$ holds from Lemma~\ref{lem2.1}, we can apply Theorem 7.1 and 
	Theorem 7.2 of \cite{ded}.  If we consider assumptions of this theorem, then the central limit theorem holds for $Z_{ni}$, therefore 
	$\mathbf{d'B}_{n0} = \frac{1}{\sqrt{n}} \sum_{i=1}^n Z_{ni}$ converges in distribution to $N(0, \theta_d^2)$. 
	Hence $\sqrt{nh}\left\{ \hat{\theta}(t)-\theta{t}-\frac{h^2}{2}\mu_2\theta^{(2)}(t)+o(h^2)\right\}$ converges in distribution to 
	$N(0, \Sigma_{\theta})$, where covariance matrix $\Sigma_{\theta} = v_0A^{-1}\Sigma_0(A^{-1})'$. $\Box$
	
	%
	%
	%
	%

	\section{Conclusions}
	In this work we derived a general seasonal time series model with  $k$-dependent and $\lambda$-dependent errors, 
	which are new concepts of dependence. In this model we derived the consistency and asymptotic normality 
	of non-parametric estimates constructed by local linear method. \\

	{\bf Acknowledgement}. We would like to thank anonymous referees for their valuable comments and suggestions. 
	This work was supported in UNESCO/China (The Great Wall) Co-Sponsored Fellowships Programme(2010).


\begin{thebibliography}{20}
	
	\bibitem{cai1}
	Z.W. Cai, R. Chen, Flexible Seasonal Time Series Models, Advances in Econometrics, 
	Econometric Analysis of Financial and Economic Time series/part B, 20(2006), 63-87.
	
	\bibitem{cai2}
	 Z.W. Cai, Trending time varying coefficient time series models with serially correlated errors, 
	J. Econometrics, 136(2007), 163-188.
	
	\bibitem{cai3}
	 Z.W. Cai, Q. Li, J.Y. Park, Functional-coefficient models for nonstationary time series data, 
	J. Econometrics. 148(2009), 101-113.

	\bibitem{cai4}
	Z.W. Cai, X. Wang, Nonparametric estimation of conditional VaR and expected shortfall, 
	J. Econometrics. 147(2008), 120-130.

	\bibitem{bar}
	 J.M. Bardet, P. Dokhan, G. Lang, N. Ragache, Depnedent Lindeberg central limit theorem and some applications, 
	arXiv: 0701872v1.
	
	\bibitem{bul}
	A.V. Bulinski, A.P. Sashkin. Strong invariance principle for dependent multi-indexed random variables, Doklady. Mathematics, 
	   72-11(2005), 503-506, MAIK Nayka/Interperiodica. 325–363.
	
	\bibitem{ded}
	J. Dedecker, P. Doukhan, G. Lang, J. R. Leon, S. Louhichi, C. Prieur, Weak dependence: with examples and applications, 
	Springer, 2007.

	\bibitem{kim}
	Non-stationary structural model with time-varying demand elasticities, J. Statist. Plann. Inference. 140(2010), 3809-3819.
	
	\end{thebibliography}
 	\end{document}